\documentclass[aps,prl,reprint,groupedaddress]{revtex4-2}

\usepackage{graphicx}

\begin{document}

\title{Role of the constriction angle on the clogging by bridging of suspensions of particles}

\author{Nathan Vani,$^{1,2}$ Sacha Escudier,$^{1}$ Deok-Hoon Jeong,$^{1}$ Alban Sauret$^{1}$}
\email[]{asauret@ucsb.edu}

\affiliation{$^{1}$ Department of Mechanical Engineering, University of California, Santa Barbara,
California 93106, USA \\
$^{2}$ Laboratoire de Physique et Mécanique des Milieux Hétérogènes (PMMH), CNRS UMR-7636, ESPCI, PSL University, Paris 75005, France}

\date{\today}

\begin{abstract}
Confined flows of particles can lead to clogging, and therefore failure, of various fluidic systems across many applications. As a result, design guidelines need to be developed to ensure that clogging is prevented or at least delayed. In this Letter, we investigate the influence of the angle of reduction in the cross-section of the channel on the bridging of semi-dilute and dense non-Brownian suspensions of spherical particles. We observe a decrease of the clogging probability with the reduction of the constriction angle. This effect is more pronounced for dense suspensions close to the maximum packing fraction where particles are in contact in contrast to semi-dilute suspensions. We rationalize this difference in terms of arch selection. We describe the role of the constriction angle and the flow profile, providing insights into the distinct behavior of semi-dilute and dense suspensions.
\end{abstract}

\maketitle

Clogging is a major issue in a wide range of systems at various length and time scales \cite{patnaik1994vascular, liu2019particle, piton2022debris,dincau2023clogging}. Clogging can happen when a flowing suspension is geometrically confined, particularly in the presence of constrictions, and can impair many applications. Therefore, guidelines to design resilient fluidic systems that handle suspensions must be carefully established. Describing clogging is challenging as it might occur through different mechanisms: sieving \cite{sauret2014clogging,sauret2018growth,duchene2020clogging,majekodunmi2022flow} bridging \cite{marin2018clogging,vani_influence_2022} and aggregation\cite{dersoir2015clogging,gerber2018particle,dersoir2019dynamics,dincau2022clog}, or a combination of thereof \cite{dressaire2017clogging}. We focus here on bridging, corresponding to the formation of a stable arch of several particles at a constriction. Recent studies have shown that semi-dilute and dense suspensions follow a general behavior for clogging similar to dry grains in silos despite some important differences (\textit{e.g.}, \cite{goldsztein2004suspension, koivisto2017effect, marin2018clogging}). More generally, the role of various parameters on bridging was investigated and included the constriction width \cite{valdes2006particle, guariguata2012jamming, marin2018clogging}, particle roughness \cite{hsu2021roughness}, particle stiffness \cite{bielinski2021squeezing}, liquid driving force \cite{souzy2022role}, as well as active particles \cite{al_alam_active_2022}. Geometrically, the ratio of the constriction width $W$ to the particle diameter $d$ is key to determining the probability of clogging \cite{dressaire2017clogging}. Experiments with dry grains in a 2D hopper suggest that the constriction angle $\theta$ could also affect the clogging probability of suspensions through the selection of the minimum possible number of particles to form an arch \cite{zuriguel2014invited}. An example of the importance of this angle might be found in additive manufacturing, as recent research has shown that better design of the nozzle angle can prevent the clogging of fiber-filled polymer ink in 3D printers \cite{croom2021mechanics}.

The difference between semi-dilute and dense suspensions has been studied through the evolution of the clogging probability with the particle fraction, revealing fundamental differences in the clogging of semi-dilute and dense suspensions \cite{vani_influence_2022}. In the former, the number of particles forming an arch is minimized and equals $N_{\rm p} = \lfloor {W}/{d} \rfloor+1$, \textit{i.e.}, the minimum size of an arch spanning the entire constriction. In contrast, dense suspensions form arches with a variety of numbers of particles $N_{\rm p} \geq \lfloor {W}/{d} \rfloor+1$. This difference in arch formation leads to a step-wise evolution of the clogging time with $W/d$ for semi-dilute suspensions and a continuous monotonic increase of the clogging time with $W/d$ for dense suspensions \cite{vani_influence_2022}, similar to the behavior of dry granular material in 2D and 3D hoppers \cite{janda2008jamming,zuriguel2005jamming,zuriguel2014invited,thomas2015fraction, nicolas2018trap,gella2018decoupling}. Since the selection of the number of particles in the arch is critical in the reduction of the clogging probability of dry grains in silos \cite{lopez2019effect}, it is unclear if tuning the angle of a constriction of a fluidic system will lead to the same reduction of clogging probability and therefore if this approach is, at all, useful for suspensions of particles.

In this Letter, we investigate the influence of the constriction angle $\theta$ on the clogging of millifluidic channels by non-Brownian monodisperse suspensions. This study is geometrically bounded by two cases: a hopper, typical of dry granular flows, with a constriction angle $\theta=90^{\rm o}$ so that the reduction from the channel to the constriction is abrupt, and a tapered long channel at small values of $\theta$ when the reduction is smooth. For dry granular materials, the seminal work of To \textit{et al.} showed that the angle of the constriction has little impact on clogging for hoppers above a critical angle \cite{to2001jamming}. This result was experimentally extended and further characterized by López-Rodríguez \textit{et al.} \cite{lopez2019effect}, and confirmed in other configurations \cite{yu2021clogging, khalid2021study}. However, the influence of the constriction angle remains elusive in the case of suspensions \cite{mondal2016coupled}. In the following, we characterize the clogging probability from hopper-like geometries to gently tapered channels for both semi-dilute and dense suspensions in viscous flows.

\begin{figure}
\centering
\includegraphics[width = 0.4\textwidth]{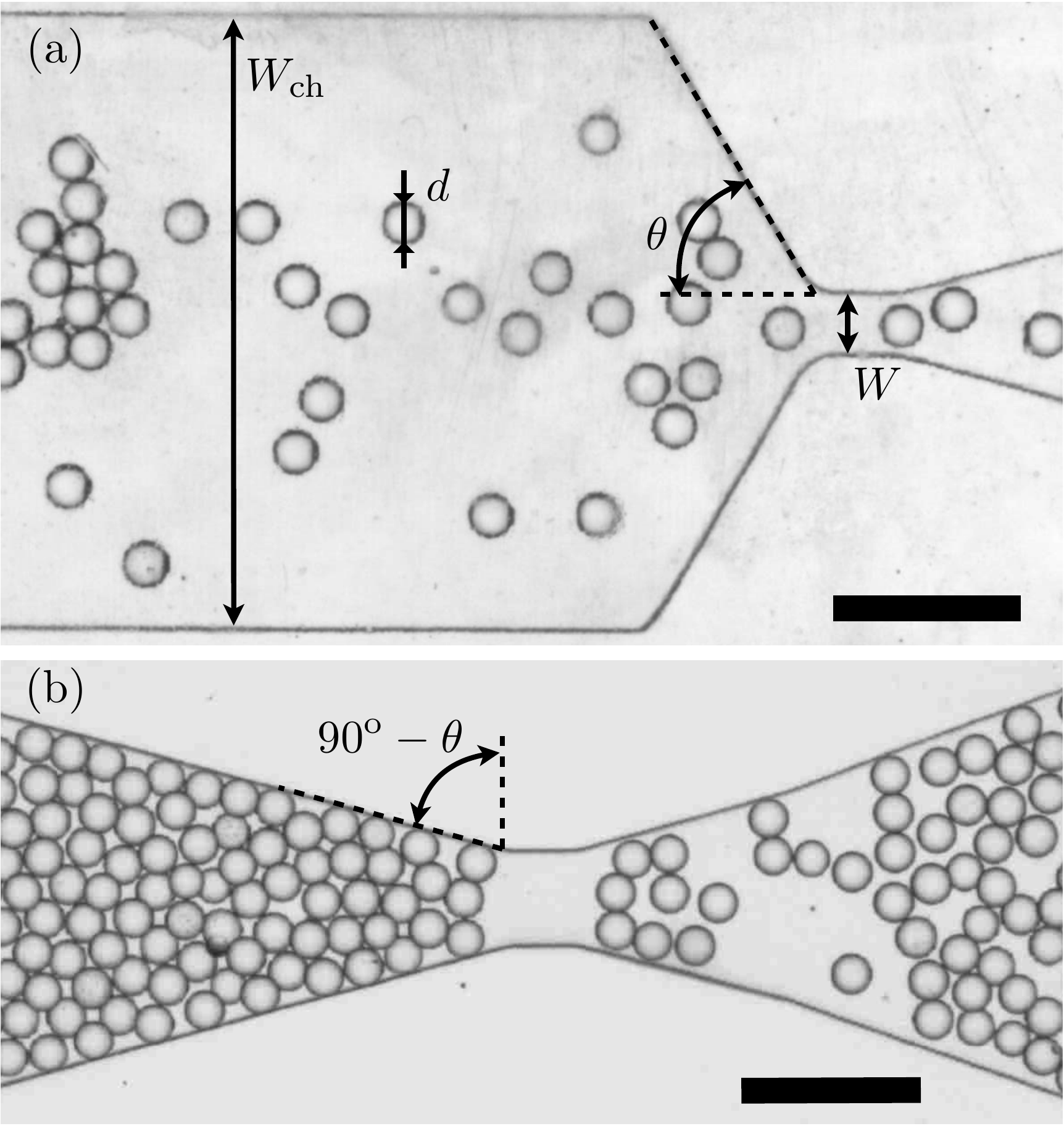}
\caption{Examples of visualizations for (a) a large constriction angle ($\theta=60^{\rm o}$ and $W/d=1.7$) with a semi-dilute suspension at $\phi=0.19$, and (b) a smaller constriction angle ($\theta=15^{\rm o}$ and $W/d=2.7$) with a dense suspension at $\phi \simeq \phi_{\rm m}$. The particles of diameter $d$ are flowing from left to right, first in a main channel of width $W_{\rm ch}=7.8\,{\rm mm}$, and then brought to a constriction of width $W$ with a narrowing part of constant angle $\theta$. Scale bars are 3 mm.} 
\label{fig:Figure1_Setup}
\end{figure}

The flow and clogging of non-Brownian suspensions are investigated in 3D-printed millifluidic channels. We can track and count the particles as well as characterize the appearance and geometry of clogs formed. The millifluidic devices are made using stereo-lithography (SLA) printing methods (Formlabs Form 3) \cite{vani_influence_2022}. The system is quasi-bidimensional with a height $H \simeq 0.95\,{\rm mm}$, minimizing particles overlap. The design of the system is illustrated in Figs. \ref{fig:Figure1_Setup}(a)-(b) and allows us to tune the width and the angle of the constriction, respectively $W$ and $\theta$. The width of the channel upstream of the constriction $W_{\rm ch}$ is kept constant in this study.

The suspensions consist of non-Brownian spherical polystyrene particles of diameter $d=580\pm15\,\mu{\rm m}$ (Dynoseeds from Microbeads) dispersed in a Newtonian mixture of DI water and polyethylene glycol (PEG, Sigma Aldrich) at 62\%/38\% per weight, with a dynamic viscosity of $\eta_{\rm f} = 75\,{\rm mPa.s}$. The density of the interstitial fluid matches the density of the particles, $\rho_{\rm f} \simeq \rho_{\rm p} \simeq 1.05\,{\rm g.cm^{-3}}$. The size of the particles ensures that electrostatic forces are negligible with respect to contact and hydrodynamic forces and that clogging occurs only through the formation of an arch at the constriction. Considering the quasi-bidimensional situation, we define here the surface fraction of particles $\phi$ as the ratio of the projected area of particles to the total area \cite{vani_influence_2022}. In the following, we investigate the influence of the angle of the constriction $\theta$ on the clogging probability. We consider both the cases of semi-dilute ($\phi \sim 0.19$ and $0.32$) and dense ($\phi \sim \phi_m$) suspensions. The suspension is injected at a constant flow rate $Q=1\,{\rm mL/min}$ leading to particle velocity of order $u_{\rm p} \sim 2.2 {\rm mm.s^{-1}}$ upstream of the constriction and a particle Reynolds number $Re_{\rm p}=\rho_{\rm f}\,du_{\rm p}/\eta_{\rm f}$ always smaller than unity.

\begin{figure}
\centering
\includegraphics[width = 0.5\textwidth]{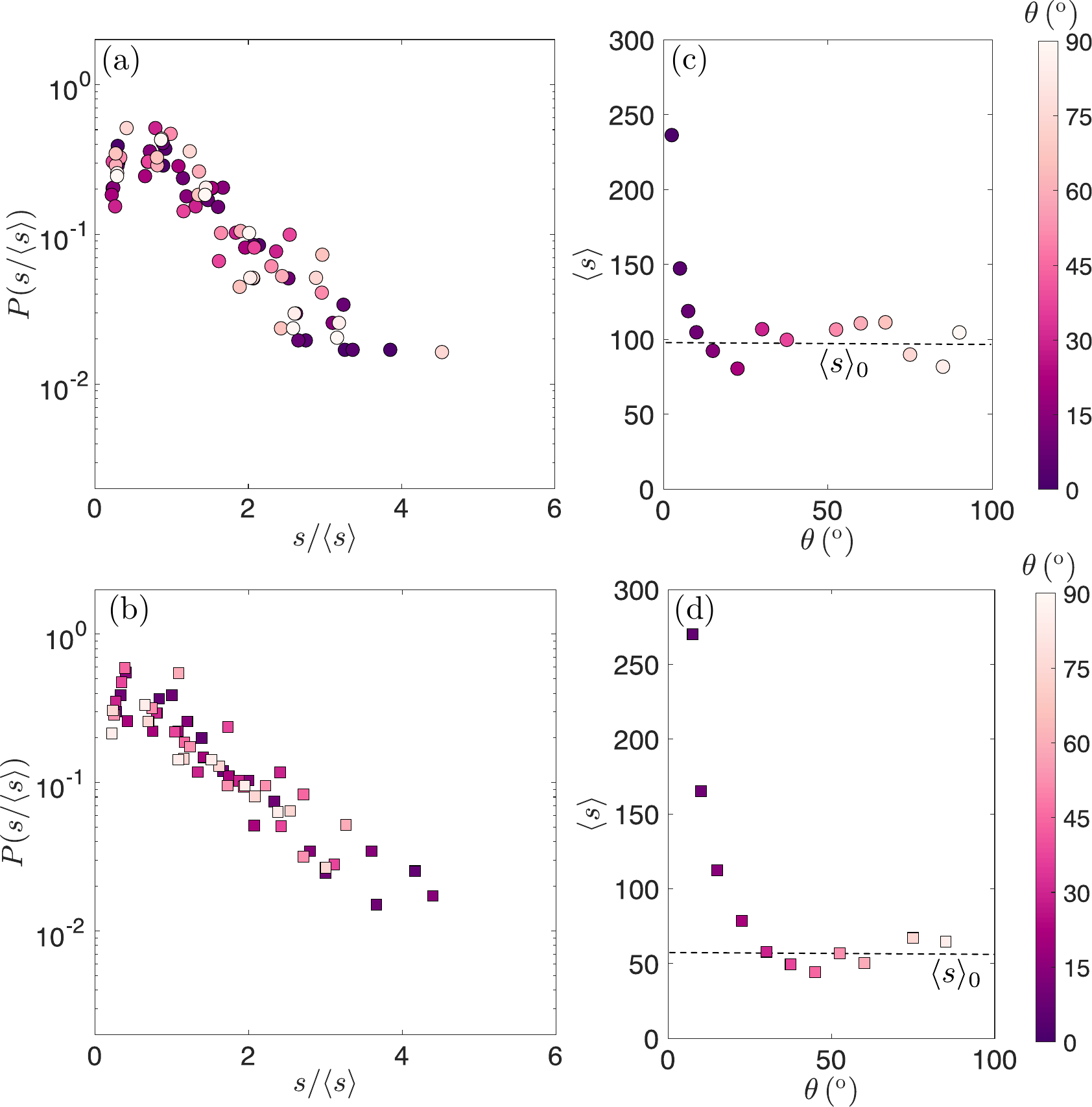}
\caption{Probability density function $P(s/\langle s \rangle)$ of the normalized number of particles escaping the constriction $s/\langle s \rangle$ before clogging for (a) $\phi=19\%$ with a constriction width $W/d=1.7$ and (c) $\phi=\phi_{\rm m}$ with a constriction width $W/d=2.7$. Evolution of the average number of particles flowing through the constriction $\langle s \rangle$ as a function of the constriction angle $\theta$ for (b) $\phi=19\%$ and $W/d=1.7$ and (d) $\phi=\phi_{\rm m}$ with $W/d=2.7$. The colorbar on the right indicates the angle of the constriction.} 
\label{fig:Figure_2_Histogram}
\end{figure}

We first performed the experiments with a semi-dilute suspension at $\phi=0.19$ for constriction angles $2.5^{\rm o}\leq \theta \leq 90^{\rm o}$. Similar to previous studies for semi-dilute suspensions \cite{marin2018clogging,vani_influence_2022}, the present experiments show that the number of particles flowing through the constriction before clogging is characterized by an exponential distribution, $p(s) \propto {\rm e}^{-s}$. In addition, using the mean number of escapees $\langle s \rangle$, the experimental data for all constriction angles collapse on a master curve, as reported in Fig. \ref{fig:Figure_2_Histogram}(a). This observation is expected since the clogging process is stochastic and can thus be described by a geometric law \cite{to2001jamming}. Similarly, a dense suspension at $\phi \sim \phi_{\rm m}$ also exhibits a similar behavior once rescaled by $\langle s \rangle$ [Fig. \ref{fig:Figure_2_Histogram}(b)]. We thereafter rely on $\langle s \rangle$ to characterize the system while varying the different parameters, and in particular the constriction angle.

We show in Figs. \ref{fig:Figure_2_Histogram}(c) and Fig. \ref{fig:Figure_2_Histogram}(d) the evolution of $\langle s \rangle$ for a semi-dilute and a dense suspension, respectively, as a function of the constriction angle $\theta$. Due to the significant experimental time to clog a constriction at low volume fractions, we used in Fig. \ref{fig:Figure_2_Histogram}(b) a constriction size of $W/d=1.7$, whereas we were able to consider $W/d=2.7$ for the dense case. The behaviors observed appear qualitatively similar with a substantial increase of $\langle s \rangle$ at small constriction angles and an almost constant value of $\langle s \rangle $ for angle larger than $\theta =10-20^{\rm o}$. However, the relative increase of $\langle s \rangle$ seems more limited for the semi-dilute suspension than for the dense suspension.

\begin{figure}
\centering
\includegraphics[width = 0.45\textwidth]{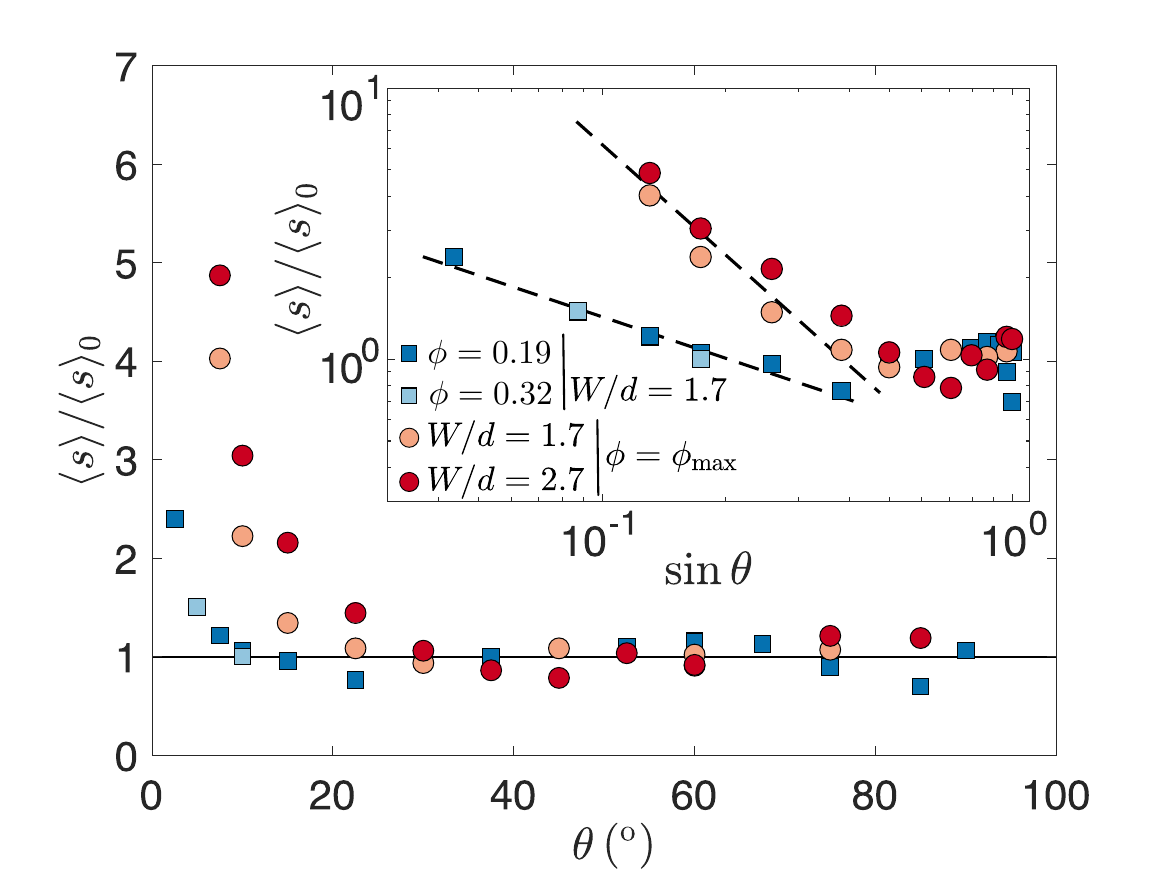}
\caption{Evolution of the rescaled average number of particles flowing through the constriction before clogging $\langle s \rangle/\langle s \rangle_0$ as a function of the angle of the constriction for semi-dilute and dense suspensions. Inset: Divergence of $\langle s \rangle/\langle s \rangle_0$ in the limit of small angles $\theta$.}
\label{fig:Figure3_All_s}
\end{figure}

To characterize more quantitatively the relative influence of the constriction angle in the semi-dilute and the dense case, we rescale the mean number of escapees, $\langle s \rangle$, by the plateau value observed for $\theta  \gtrsim 30^{\rm o}$, later referred to as $\langle s \rangle_0$ and shown by the horizontal dotted line in Fig. \ref{fig:Figure_2_Histogram}(c) and \ref{fig:Figure_2_Histogram}(d).

Fig. \ref{fig:Figure3_All_s} shows that reducing the constriction angle has indeed a stronger relative effect for dense suspensions where particles are more in contact. In our system, for an angle of $\theta=5^{\rm o}$, the millifluidic channel can take as long as 7 times the time it would take for a dense suspension to clog at a larger constriction angle. The role of the volume fraction $\phi$ and the increase in $\langle s \rangle/\langle s \rangle_0$ at small constriction angle can be better seen in the inset of Fig. \ref{fig:Figure3_All_s}. In all cases, the experimental results suggest that $\langle s \rangle/\langle s \rangle_0 \propto (\sin \theta)^\alpha$ where $\alpha<0$ depends on the experimental condition. This scaling law observed comes from the fact that the reaction of the particles with the solid surface is proportional to $\sin \theta$. In particular, for semi-dilute suspension and a small constriction $W/d=1.7$ we observe that for both $\phi=0.19$ and $\phi=0.32$, $|\alpha| < 1$ ($|\alpha|=0.5 \pm 0.1$). However, for dense suspensions ($\phi \sim \phi_{\rm m}$) the exponent is larger, $|\alpha|=1.1 \pm 0.3$, and seems similar for the two dimensionless constriction width $W/d$ considered here, although the data suggests that for $W/d=1.7$ the coefficient is closer to $|\alpha|=0.9$ and for $W/d=2.7$ is closer to $|\alpha|=1.2$. This observation suggests that reducing the angle of the constriction is more efficient in delaying clogging for concentrated suspensions with particles almost always in contact.

Although their focus was more on the evolution of clogging with $W/d$ rather than the angle, we can consider the experimental data of L\'opez-Rodr\'iguez \textit{et al.} \cite{lopez2019effect} for some values of $W/d$ (see supplementary material). In their experiments, an effect of the angle seems to appear as early as $\theta=30^{\rm 0}$ (note that the definition of the angle of the constriction is different, and we adapted their data to our notation). In addition, the increase of $\langle s \rangle/\langle s \rangle_0$ when reducing the constriction angle is even more pronounced. A fit to their data suggests that $\langle s \rangle/\langle s \rangle_0 \propto (\sin \theta)^\alpha$ still apply. However, the value obtained from their experiments at a similar $W/d$ than our dense suspension is $| \alpha | \simeq 2.6$, which is significantly larger than the one observed for dense suspensions. Moreover, in their data the value of $| \alpha |$ varies based on $\lfloor W/d \rfloor$ (see supplementary material). Such an effect was less significant for dense suspensions, likely due to limitations in our range of $\lfloor W/d \rfloor$. The difference in this divergence may be due to the difference in roughness and the lubrication between particles in our cases.

\begin{figure}
\centering
\includegraphics[width = 0.5\textwidth]{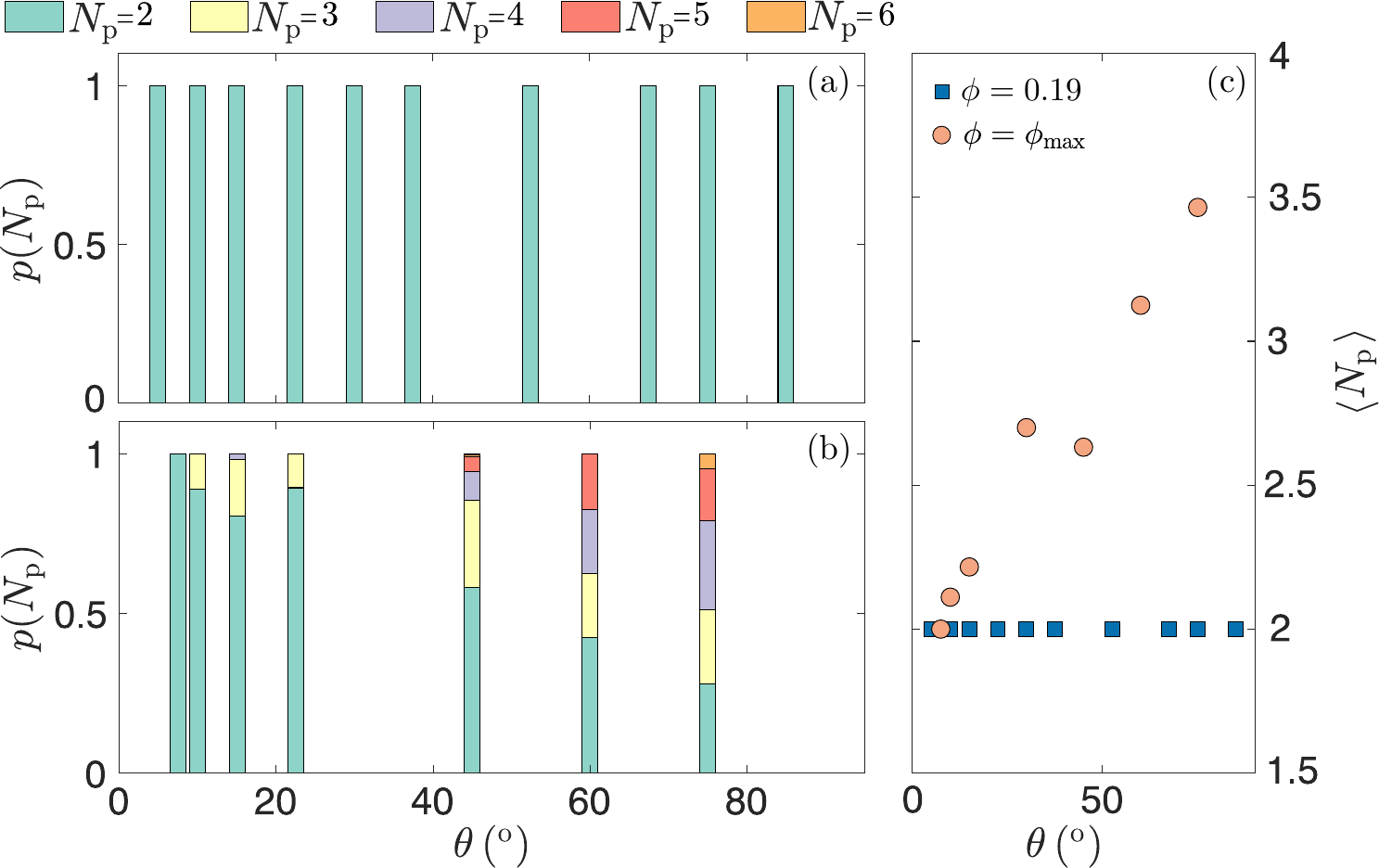}
\caption{Relative percentage of clogging arches with $N_{\rm p}$ particles when varying the angle of the constriction of width $W/d=1.7$ for (a) $\phi=0.19$ and (b) $\phi \sim \phi_{\rm m}$. The color code indicates the number of particles in the arches. (c) Evolution of the average number of particles $\langle N_{\rm p} \rangle$ in the clogging arches in the same two cases.}
\label{fig:Figure4_Arches}
\end{figure}

To rationalize the difference in exponent for $\langle s \rangle/\langle s \rangle_0 \propto (\sin \theta)^\alpha$ in the semi-dilute and dense cases, we analyzed the arches formed for all experiments, and in particular the number of particles that constitute them. We report in Figs. \ref{fig:Figure4_Arches}(a) and \ref{fig:Figure4_Arches}(b) the evolution of the percentage of arches made of $N_{\rm p}$ particles for $\phi=0.19$ and $\phi_{\rm m}$, respectively. We observe that for semi-dilute suspensions, the number of particles constituting an arch for $W/d=1.7$ remains equal to $\lfloor W/d \rfloor+1=2$ irrespective of the constriction angles.  In contrast, Fig. \ref{fig:Figure4_Arches}(b) illustrates that for $\phi \sim \phi_{\rm m}$, a range of arch sizes is possible when the constriction angle is large, approximately $\theta \gtrsim 45^{\rm o}$. Arches comprised of $N_{\rm p}=2$ particles occur frequently, sometimes even more so than the minimum possible size of $N_{\rm p}=2$. However, as the constriction angle $\theta$ decreases, a trend emerges where only one arch size, equivalent to $\lfloor W/d \rfloor +1=2$, is observed, as can be seen for instance for $\theta=7.5^{\rm o}$ here. This shift is further evident in Fig. \ref{fig:Figure4_Arches}(c) when directly comparing the evolution of the average arch sizes for $\phi=0.19$ and $\phi \sim \phi_{\rm m}$. We observe a reduction of arch size along with the reduction of the constriction angle for dense suspensions, with less stable arch configurations available for clogging. This result is also valid for larger $W/d$ with dense suspensions (see Supplementary Materials). The present observation is in agreement with the one reported for granular flows \cite{to2001jamming, lopez2019effect}. 

As a result, since the reduction in arch size is not possible for semi-dilute suspensions as the arch size is already minimized, the effect of decreasing the constriction angle is less pronounced for semi-dilute suspensions. This is the main cause behind the distinct divergent behavior of semi-dilute and dense suspensions. Yet, even without this selection mechanism, the mechanical stability of the arches decreases when decreasing the angle, leading to less stable arches and thus an increase in $\langle s \rangle$.

\begin{figure}
\centering
\includegraphics[width = 0.5\textwidth]{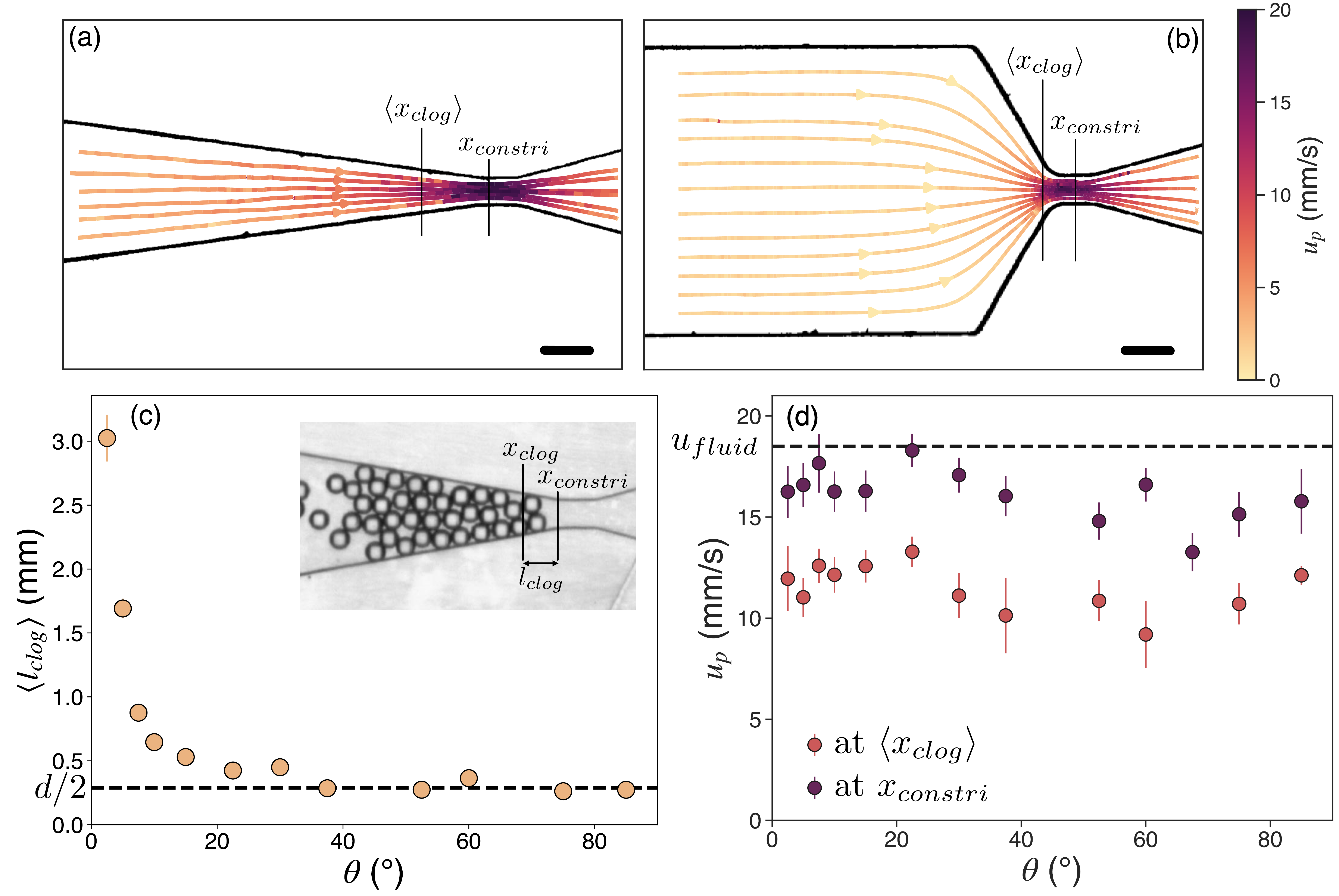}
\caption{Example of streamlines and velocity amplitude (color bar) for $W/d=1.7$, $\phi=0.19$ and (a) $\theta=7.5^{\rm o}$ and (b) $\theta=60^{\rm o}$. Scale bars are 2 mm. (c) Average distance between the position of a clog and the constriction as a function of the constriction angle. (d) Velocities of the particles when crossing the constriction at its narrowest part $x_{constri}$ and at the average position of clogs $\langle x_{clog} \rangle$ as a function of the constriction angle $\theta$ for $\phi=0.19$. The horizontal dashed line is the velocity of the fluid computed with the imposed flow rate.}
\label{fig:Figure5_Hydrodynamic}
\end{figure}

In the case of semi-dilute suspensions, fluid flow may affect the clogging dynamics and stability of the arches. We report the trajectory of the particles in Fig. \ref{fig:Figure5_Hydrodynamic}(a) for $\theta=7.5^{\rm o}$ and \ref{fig:Figure5_Hydrodynamic}(b) for $\theta=60^{\rm o}$. The trajectories of the particles follow qualitatively the fluid streamlines, but particles have a slightly lower velocity due to the vertical confinement. Since incoming particles will impact on arches just after their formation occuring at different locations depending on the angle, we consider in Fig. \ref{fig:Figure5_Hydrodynamic}(c) the average distance between the position of a clog (defined at the most upstream edge of the arch) and the constriction, $\langle l_{\rm clog}\rangle$. As expected $\langle l_{\rm clog}\rangle$ increases when decreasing $\theta$ since the clogs form earlier in the channel. We can then compare in Fig. \ref{fig:Figure5_Hydrodynamic}(d) the velocity of particles when entering in contact with a clog and at the constriction for different angles. Both velocities keep a similar value regardless of the constriction angle. While this observation was expected at the constriction, it is also the case at the average position of clogs as this location moves further away from the constriction when decreasing the angle [Fig. \ref{fig:Figure5_Hydrodynamic}(c)]. Therefore, stable arches are created, on average, at the part of the constriction with the same velocity. Therefore, we do not expect arch destabilization due to the impact of particles to be more important at small constriction angles for viscous flows. Another main difference between low and large angles is the dispersion of the streamlines. As particles arrive in the tapered part of the channels, the more confined environment will lead to more interaction and more rearrangements before the constriction. Such effects might lead to a smoother flow rate of particles through the constriction and, thus, to fewer variations in the local solid fraction directly upstream of the constriction. Nonetheless, our experiments seem to suggest that hydrodynamics do not appear to play the main role in the decrease of the clogging probability as it is mostly governed by particle-channel interactions, at least for small enough constrictions, $W/d < 2$ here.

In conclusion, our work characterized the role of the constriction angle on the bridging of semi-dilute and dense suspensions. The angle of constriction has little to no impact when $\theta \gtrsim 20^{\rm o}$. This value angle is likely to depend on the frictional interactions and is similar to dry grains. Our results revealed that, while both the divergence at low angles for semi-dilute and dense suspensions scales as $\left( \sin\theta\right)^\alpha$ ($\alpha<0$), the latter is far more sensitive to it. This key difference comes from the fact that the average number of particles in an arch is correlated to the constriction angle for dense suspension. Decreasing $\theta$ in this case, eliminates the possibility of forming arches with more than $N_{\rm p} = \lfloor {W}/{d} \rfloor+1$ particles due to geometrical constraints, as reported for dry grains in silos \cite{lopez2019effect} and thus reduce significantly the clogging probability. The role of $\theta$ is less important for semi-dilute suspensions since arches are already formed with the minimal number of particles $N_{\rm p} = \lfloor {W}/{d} \rfloor+1$ regardless of $\theta$. The smaller divergence observed is due to the mechanical stability of the arch, an effect that is present in both cases. Finally, while hydrodynamics may play a role in the clogging of semi-dilute suspensions, it does not seem to be the main mechanism for the reduction of the clogging probability when reducing the constriction angle in our experiments. Numerical simulations that would allow computing the drag force on the particles and the particle-particle and particle-wall interactions could help decipher more accurately the contribution of both mechanisms \cite{zhou2024simplified}.

\begin{acknowledgments}
This material is based upon work supported by the National Science Foundation under NSF
Faculty Early Career Development (CAREER) Program Award CBET No. 1944844.
\end{acknowledgments}

\bibliographystyle{apsrev4-1}
\bibliography{Biblio_Clogging}

\newpage

\onecolumngrid
\appendix

\begin{center}
\textbf{\Large Supporting Material for} 
\medskip

\noindent \textbf{\large Role of the constriction angle on the clogging by bridging of suspensions}
\end{center}

\vspace{7mm}

\noindent {\Large {\textsc{Absence of correlation between successive clogging events}}}

\vspace{3mm}

\noindent Two clogging events by bridging of cohesionless non-Brownian particles should be independent of each other. In our experiments, for all angles considered, series of trials were conducted in a sequential process. For semi-dilute suspensions, in the event of a clog, the device was flushed with fluid alone, eliminating any particles within the system. Therefore, each trial starts with a new batch of suspensions, allowing for a precise control of the solid fraction. For dense suspensions, whenever a clog occurred, the flow direction was reverted for a sufficiently long time to break the clog and ensure that we reinitialized the system. The flow direction was then reversed again to return to the initial conditions. This cycle was repeated a sufficient number of times for every set of parameters until we obtained statistically relevant results. 

\vspace{2mm}

\begin{figure}[h!]
    \centering
    \includegraphics[width=0.75\textwidth]{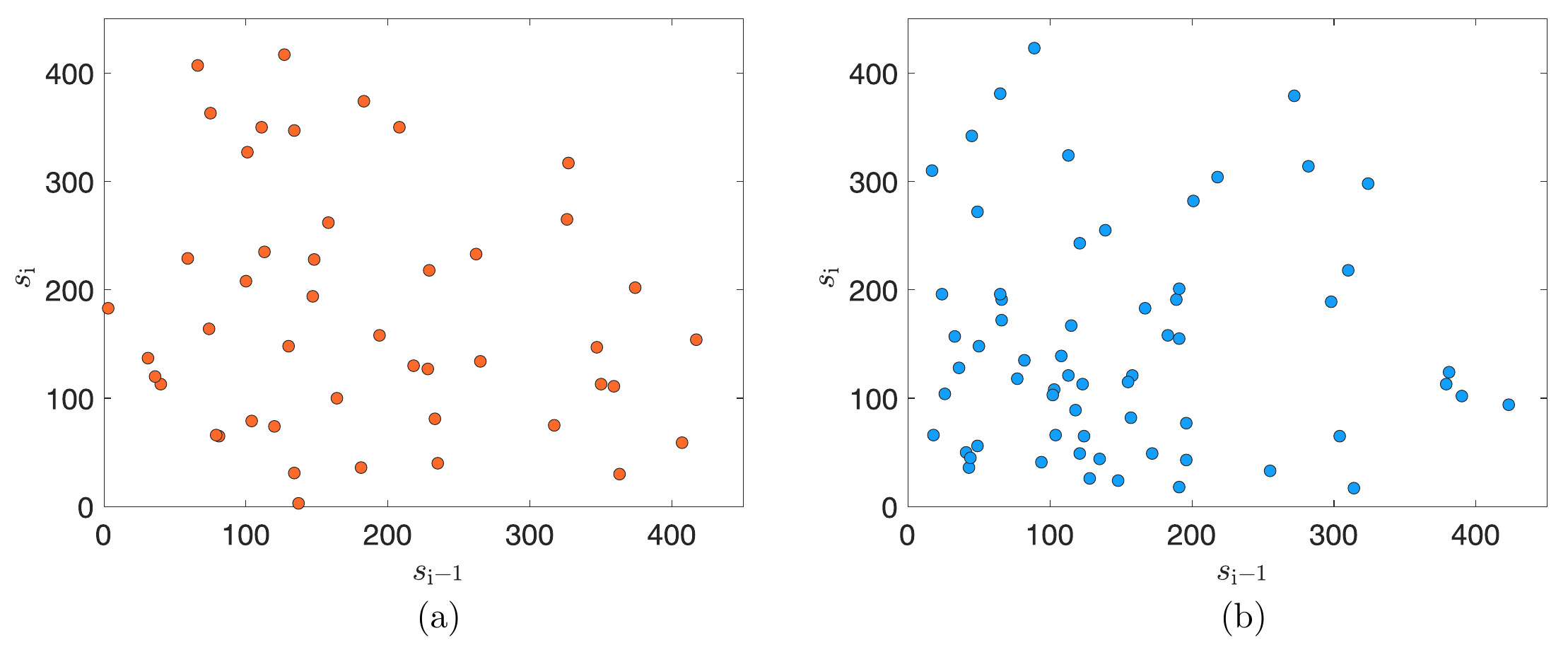}
    \caption{{\small{Number of particles flowing through the constriction before a clogging event occurs for the experiments $i$, $s_i$, as a function of the number of particles in the previous clogging event, $s_{i-1}$, for (a) a semi-dilute suspension ($\phi \simeq 0.19$, $W/D=1.7$, $\theta =2.5^{\rm o}$) and (b) a dense suspension ($\phi \simeq \phi_{\rm m}$, $W/d = 2.7$, $\theta=10^{\rm o}$).}}}
    \label{fig:SM_1}
\end{figure}

\noindent We report in Fig. \ref{fig:SM_1}(a) and Fig. \ref{fig:SM_1}(b) the number of escapees for a clogging event $i$ as a function of the number of escapees in the previous clogging event $i-1$, for semi-dilute and dense suspensions, respectively. The dispersion of the data indicates that our experimental methods do not lead to any apparent correlation between consecutive events, \textit{i.e.}, the history of the system does not influence subsequent clogging events. 

\vspace{10mm}

\noindent {\Large {\textsc{Evolution of $\langle s \rangle/\langle s \rangle_0$ with the hopper angle for dry grains}}}

\vspace{3mm}

\begin{figure}[h!]
    \centering
    \includegraphics[width=0.5\textwidth]{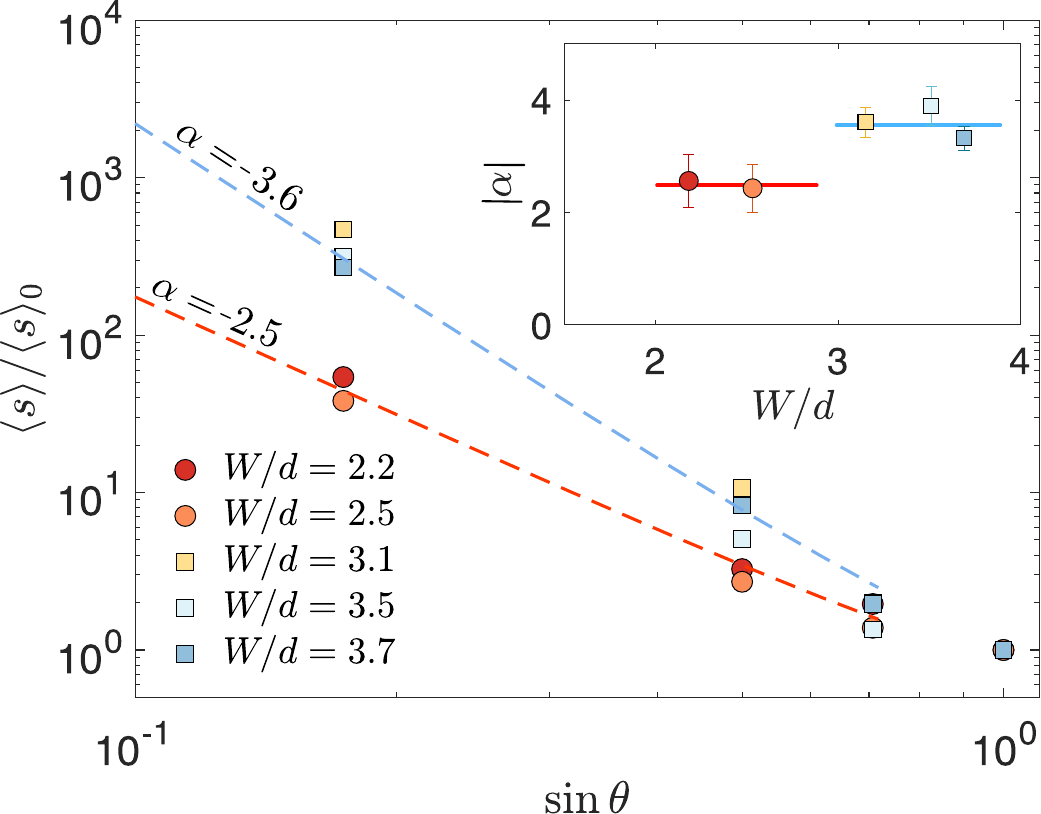}
    \caption{{\small{Evolution of the rescaled average number of particles flowing through the constriction before clogging, $\langle s \rangle/\langle s \rangle_0$, as a function of the angle of the constriction of the hopper, $\theta$. Data are extracted from the experiments of L\'opez-Rodr\'iguez \textit{et al.} \cite{lopez2019effect}. Inset: Exponent $|\alpha|$ characterizing the divergence as $\langle s \rangle/\langle s \rangle_0 \propto (\sin \theta)^{\alpha}$ ($\alpha <0$) in the limit of small angles $\theta$.}}}
    \label{fig:Appendix_1}
\end{figure}

\noindent L\'opez-Rodr\'iguez \textit{et al.} \cite{lopez2019effect} have considered the role of the narrowing angle on the clogging of dry grains discharged from a two-dimensional hopper under the action of gravity. Since their system shares some common features with our millifluidic experiments, but also some key differences, we report their experimental data for different $W/d$ in Fig. \ref{fig:Appendix_1}. Note that their definition of the constriction angle differs from our definition in that it is complementary (\textit{i.e.}, a constriction angle of $60^{\rm o}$ in their study corresponds to $90^{\rm o}-60^{\rm o}=30^{\rm o}$ with our notation). L\'opez-Rodr\'iguez \textit{et al.} kept $\theta$ constant and varied $W/d$. As a result, it is not possible to get exactly the same value of $W/d$ and vary the angle so that we interpolate between appropriate values of $W/d$ to get the values presented here. We define $\langle s \rangle_0$ as the average number of escapees in the plateau region ($\theta > 40^{\rm o}$).

We report in Fig. \ref{fig:Appendix_1} the evolution of $\langle s \rangle/\langle s \rangle_0$ as a function of the angle of the constriction. Interestingly, their experimental results show a similar qualitative behavior as the one observed in our experiments. In particular, it shows a similar divergence, scaling as $\langle s \rangle/\langle s \rangle_0 \propto (\sin \theta)^\alpha$. However, $| \alpha | \simeq 2.5$ is significantly larger than the one observed for dense suspensions at a similar $W/d$ in our experiments. Moreover, in their data the value of $| \alpha |$ varies widely based on $\lfloor W/d \rfloor$. Such an effect was not observed for dense suspensions, likely due to the factor that our range of $\lfloor W/d \rfloor$ was more limited due to experimental constraints.

\vspace{10mm}

\noindent {\Large {\textsc{Size distribution of arches for dense suspensions at $W/d=2.7$}}}

\vspace{5mm}

\begin{figure}[h!]
\centering
\includegraphics[width = 0.8\textwidth]{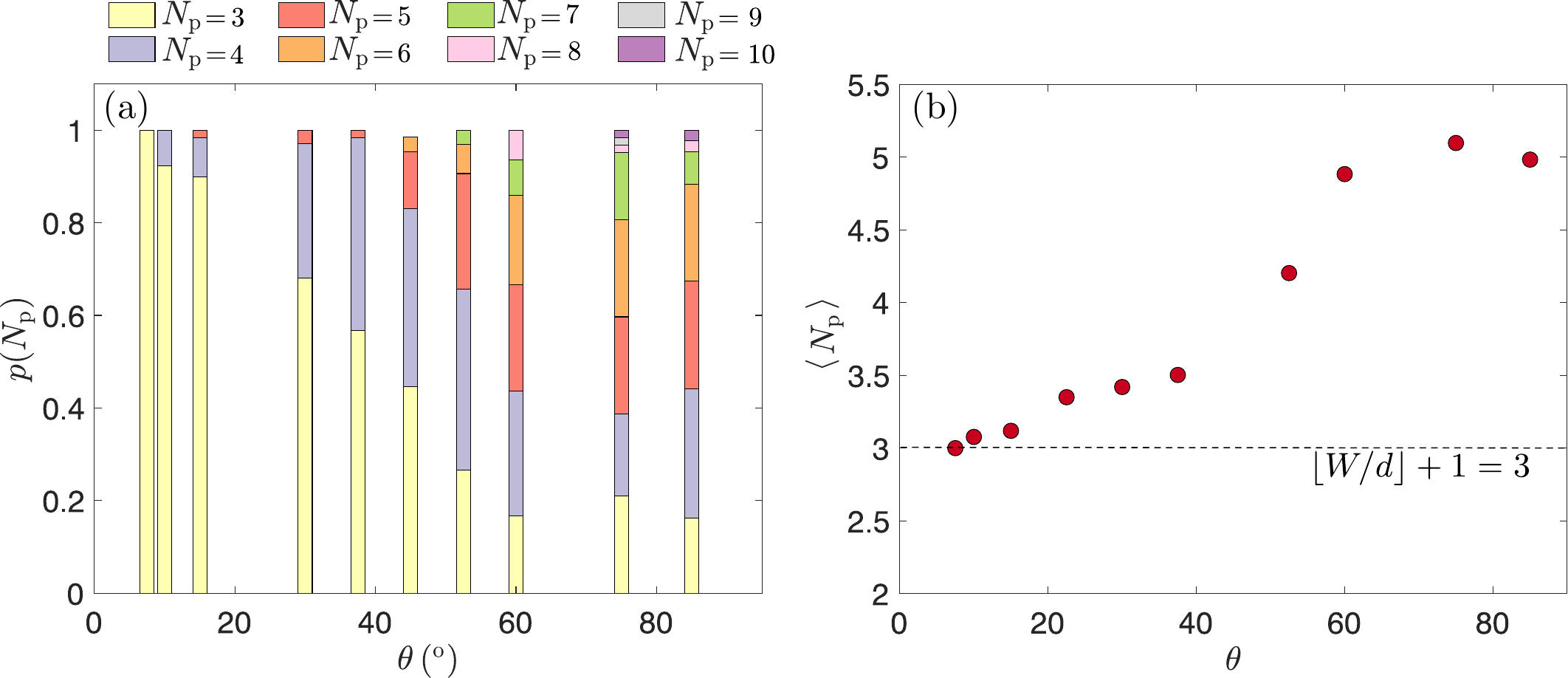}
\caption{{\small{(a) Relative percentage of clogging arches with $N_{\rm p}$ particles when varying the angle of the constriction $\theta$ for a width $W/d=2.7$ for dense suspensions ($\phi \sim \phi_{\rm m}$). The color code indicates the number of particles in the arches. (b) Evolution of the average number of particles $\langle N_{\rm p} \rangle$ in the clogging arches with the constriction angle in the same configurations.}}}
\label{fig:Appendix_2}
\end{figure}

\noindent In Figs. 4(a) and 4(b) in the main paper, we compared side by side the distribution of the number of particles in clogging arches for semi-dilute and dense suspensions, respectively, for the same constriction width, $W/d=1.7$. We report here in Fig. \ref{fig:Appendix_2}(a) the distribution $p(N_{\rm p})$, \textit{i.e.}, the number of particles that constitute the arches formed for a dense suspension ($\phi \sim \phi_{\rm m}$) and a constriction width $W/d=2.7$. We observe a similar behavior with a range of possible sizes of arches formed for dense suspension at $\theta \gtrsim 30^{\rm o}$. Decreasing the constriction angle $\theta$ here also leads to a selection of the size of the arches formed towards $N_{\rm p}=\lfloor W/d \rfloor+1=3$, as can be seen in Fig. \ref{fig:Appendix_2}(b) where the average arch size (in number of particles) $\langle N_{\rm p} \rangle$ is reported.

\medskip

\end{document}